\documentclass[twocolumn,showpacs,prl]{revtex4}
\usepackage{graphicx}
\newcommand{\x}[1]{{\textstyle #1}}
\newcommand{\xrm}[1]{\mbox{ #1}}
\newcommand{\fnd}[2]{\frac{\textstyle #1}{\textstyle #2}}
\newcommand{\bm}[1]{\mbox{\boldmath $#1$}}
\newcommand{\dissum}[2]{\displaystyle \sum_{#1}^{#2}}
\newcommand{\abs}[1]{\left| #1\right|}
\begin{document}
\title{\bm{D_{sJ}}(2860) as the first radial excitation of the
\bm{D_{s0}^{\ast}}(2317)}
\author{Eef van Beveren}
\email{eef@teor.fis.uc.pt}
\affiliation{Centro de F\'{\i}sica Te\'{o}rica,
Departamento de F\'{\i}sica, Universidade de Coimbra,
P-3004-516 Coimbra, Portugal}
\author{George Rupp}
\email{george@ist.utl.pt}
\affiliation{Centro de F\'{\i}sica das Interac\c{c}\~{o}es Fundamentais,
Instituto Superior T\'{e}cnico, Edif\'{\i}cio Ci\^{e}ncia, Piso 3,
P-1049-001 Lisboa, Portugal}
\date{\today}

\begin{abstract}
A coupled-channel model previously employed to describe
the narrow $D_{s0}^{\ast}$(2317) and broad $D_0^{\ast}$(2400)
charmed scalar mesons is generalized so as to include all ground-state
pseudoscalar-pseudoscalar and vector-vector two-meson channels.
All parameters are chosen fixed at published values,
except for the overall coupling constant,
which is fine-tuned to reproduce the $D_{s0}^{\ast}$(2317) mass.
Thus, the radial excitations $D_{s0}^{\ast}$(2850) and $D_0^{\ast}$(2740)
are predicted, both with a width of about 50 MeV.
The former state appears to correspond to the new $D_{sJ}$(2860) resonance
decaying to $DK$ announced by BABAR in the course of this work.
Also the $D_0^{\ast}$(2400) resonance is roughly reproduced,
though perhaps with a somewhat too low central resonance peak.
\end{abstract}


\maketitle

The discovery of the $D_{s0}^*(2317)$ \cite{PRL90p242001} and 
$D_{s1}(2460)$ \cite{PRD68p032002} charm-strange mesons three years ago has
triggered a strongly renewed interest in heavy-light mesons, and even meson
spectroscopy in general. Especially the $D_{s0}^*(2317)$ has given rise to many
different theoretical efforts (see Ref.~\cite{PRD74p037501} for a long though
still incomplete list of references). The reason is its surprisingly low mass,
some 170 MeV below the predictions of standard relativized constituent quark
models for the ground-state scalar $c\bar{s}$ meson (see e.g.\
Ref.~\cite{PRD32p189}), an assignment that has in the meantime been confirmed
by experiment \cite{PDG2006}. This discrepancy led several model builders to
propose alternative explanations for the $D_{s0}^*(2317)$, such as a tetraquark
or a meson molecule. However, in Ref.~\cite{PRL91p012003} we showed how the low
mass of the $D_{s0}^*(2317)$ can be quantitatively
understood by taking into account its strong coupling to the nearby $S$-wave
$DK$ channel. This explanation was later supported by Refs.~\cite{PLB601p137}
and \cite{PRD70p114013}. Similarly, we explained the $D_{s1}(2460)$ in
Ref.~\cite{EPJC32p493} via its strong coupling to the $S$-wave $D^*\!K$
threshold. The coupled-channel model employed in Ref.~\cite{PRL91p012003} had
been previously used, with essentially the same parameters, to reproduce the
$S$-wave $K\pi$ phase shifts and predict \cite{EPJC22p493} the $K_0^*(800)$
(alias $\kappa$) resonance, later confirmed by experiment \cite{PDG2006}.

Nevertheless,  no consensus has been reached so far on the $D_{s0}^*(2317)$,
in part due to the poor experimental status of the very broad
partner charm-nonstrange state, listed as $D_0^*(2400)$, but first reported
at a mass of 2308 MeV \cite{PRD69p112002} and later also at 2407 MeV
\cite{PLB586p11}. Therefore, a more detailed coupled-channel analysis of
charmed scalar mesons is very opportune, also in view of new and heavier states
that are expected to be found at B factories. Clearly, for a reliable
description of higher resonances, additional decay channels must be accounted
for. Thus, in the present Letter, we extend \cite{Madrid_talk} the model of 
Ref.~\cite{PRL91p012003} by including all lowest pseudoscalar-pseudoscalar (PP)
and vector-vector (VV) two-meson channels that couple to the scalar $c\bar{s}$
and $c\bar{n}$ ($n=u,d$) systems, in an approach very similar to
Ref.~\cite{HEPPH0606022}. In the latter paper, the coupling to all PP channels
allowed to fit the properties of the light scalar mesons $\sigma$, $\kappa$,
$a_0$(980), and $f_0$(980), such as phase shifts, line-shapes, elasticities,
and inelastic amplitudes, obtaining an overall good description of these
observables, as well as very reasonable pole positions. In the present
investigation, the inclusion of the VV channels as well is crucial to study
possible radial excitations, as in the $c\bar{n}$ and $c\bar{s}$ sectors the
lowest VV channels open at roughly 2.8 GeV and 2.9 GeV, respectively.

\begin{figure}[htbp]
\begin{center}
\begin{tabular}{c}
\includegraphics[height=220pt, angle=0]{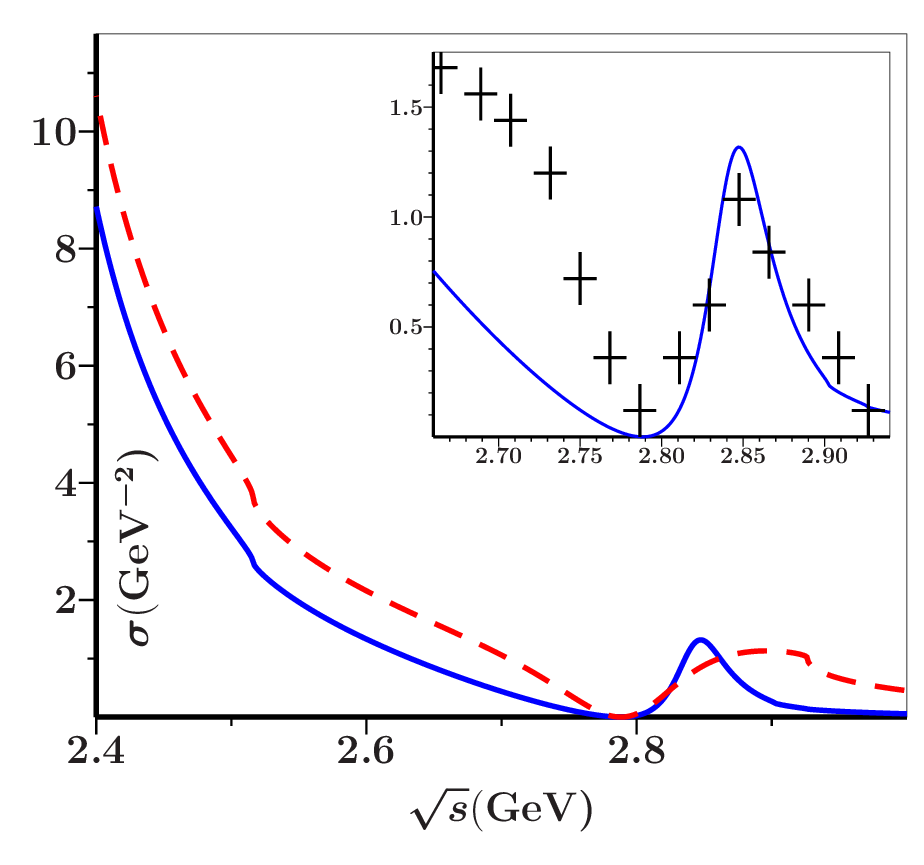}
\mbox{} \\[-20pt]
\end{tabular}
\end{center}
\caption[]{The predicted $S$-wave $DK\to DK$ cross sections.
The dashed curve corresponds to PP channels only, the solid curve
to PP + VV channels.
In the inset, we compare the model results to the data, with arbitrary
normalization taken from Ref.~\cite{DsJ2860}.}
\label{DK}
\end{figure}

We will first discuss the results and finish this Letter with a short
description of the mathematics behind the Resonance-Spectrum Expansion
(RSE), which is the framework of our model \cite{IJTPGTNO11p179}.
In Fig.~\ref{DK} we show, for the $c\bar{s}$ case,
the resulting $S$-wave $DK\to DK$ cross sections.
The dashed line refers to the case where only PP channels are included,
the solid line to the case where also VV channels are accounted for.
We will discuss the latter case.
A comparison of differences for the two situations
is presented below for the $c\bar{n}$ system.

At energies close to threshold (at 2.363 GeV),
the cross sections are large due to the presence of the
$D_{s0}^{\ast}(2317)$ bound state just below threshold.
For higher total invariant mass ($\sqrt{s}$), the cross sections
decrease, however not as fast as expected, due to the presence of
a scattering-matrix pole, which we find at $2779-i233$ MeV.
At 2.516 GeV one observes the effect of the opening of the
$D_{s}\eta$ channel, while
at about 2.79 GeV the cross sections almost vanish.
The first radial excitation of the $D_{s0}^{\ast}(2317)$
is found with a peak mass of 2847 MeV and a width
of 47 MeV, and so is a good candidate for the new BABAR state
$D_{sJ}$(2860) \cite{DsJ2860}, which decays to $DK$ and not to $D^{\ast}\!K$,
having a mass of 2857 MeV and a width of 48 MeV.
In our model it is associated with a resonance pole
at $2842-i23.6$ MeV.
From the inset of Fig.~\ref{DK},
one can judge how well our ${D_{s0}^{\ast}}'$(2850)
predicts the line shape of BABAR's $D_{sJ}$(2860).
There is furthermore some indication \cite{DsJ2860} that the data need
a broad state as well, which might correspond to our pole at $2779-i233$.

\begin{figure}[htbp]
\begin{center}
\begin{tabular}{c}
\includegraphics[height=220pt, angle=0]{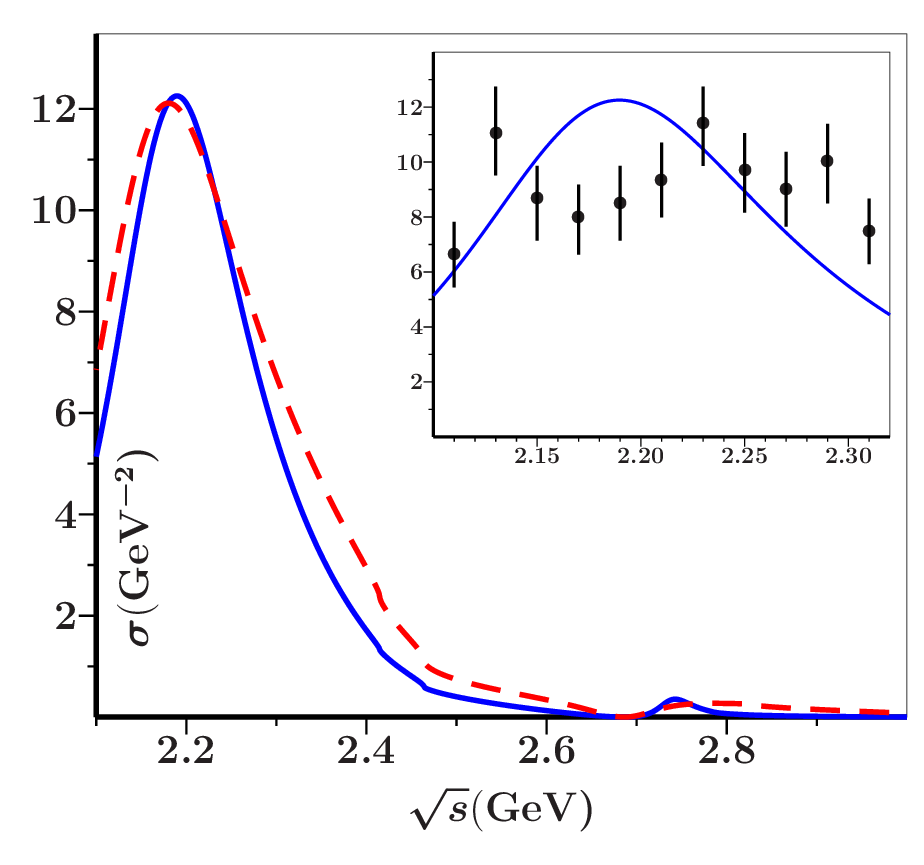}
\mbox{} \\[-20pt]
\end{tabular}
\end{center}
\caption[]{The predicted elastic $S$-wave $D\pi$ cross sections.
The dashed curve correspond to PP channels only, the solid curve
to PP + VV channels. In the inset, we compare the model results
to the data, with arbitrary normalization
taken from Ref.~\cite{PRD69p112002}.}
\label{Dpi}
\end{figure}

In Fig.~\ref{Dpi} we show, for the $c\bar{n}$ case,
the resulting $S$-wave $D\pi\to D\pi$ cross sections.
We find the lowest resonance pole at $2149-i111$ MeV (PP) or $2174-i96.4$ MeV
(PP+VV), with peak mass at 2180 MeV or 2190 MeV, respectively.
This broad resonance should correspond to the $D_0^{\ast}$(2400). 
Our prediction seems too low, but is not unreasonable in view of the unsettled
experimental situation \cite{PDG2006}, and also considering our highly
dynamical $D_0^{\ast}$(2400) pole \cite{PRL91p012003}, which can travel a long
distance with moderate changes in the model's coupling constant ($\lambda$).
For instance, if we reduce $\lambda$ somewhat so as to let the
$D_{s0}^{\ast}(2317)$ become slightly heavier, though still below the
$DK$ threshold, it is possible to increase the $D_0^{\ast}$(2400) mass
prediction by up to 100 MeV. Nevertheless, we believe it is safer to
keep the established $D_{s0}^{\ast}(2317)$ in its place, considering
the persisting uncertainties regarding the $D_0^{\ast}$(2400).
Moreover, the experimental values concern production processes, and
not elastic scattering. Furthermore, the analyses rely on a Breit-Wigner
shape for the $D^{\ast}_{2}$(2460) resonance, which has a large contribution
to the total signal. In the inset of Fig.~\ref{Dpi},
we show a comparison of our signal with the data \cite{PRD69p112002},
for invariant masses well below the $D^{\ast}_{2}$(2460) resonance.

Next we look for poles at higher energies. In this situation, we only
consider the poles for the full PP+VV system, as several VV channels
open above $\sim2.8$ GeV. Still in the $c\bar{n}$ case,
we find a relatively narrow pole at $2737-i24.0$ MeV and a very broad one
at $2703-i228$ MeV. The narrow state, with a width of about 50 MeV,
corresponds to the first radial excitation of the $c\bar{n}$ system, shifted
to complex energy by the coupled channels, while the very broad resonance
is the strongly distorted and shifted ground state of the confinement
spectrum, also found in Ref.~\cite{PRL91p012003}, though now with a width of
roughly 450 MeV instead of $\approx200$ MeV. Note that the
$D_0^{\ast}$(2400) is a dynamical continuum pole, just as in 
Ref.~\cite{PRL91p012003}. The narrow resonance at about 2.74 GeV predicted here
should be observable, though the $S$-wave elastic $D\pi$ cross section is quite
small (see Fig.~\ref{Dpi}, solid curve). 

Finally, let us turn to a short description of the model employed in this work.
The model's scattering matrix ($S$) for $N$ two-meson channels
(masses $M_{i1}$ and $M_{i2}$, $i=1$, 2, 3, $\dots$, $N$,
orbital angular momentum $\ell_{i}$, relative linear momentum $k_{i}$),
all coupled to one quark-antiquark confinement channel
(radial confinement spectrum given by
$E_{0}=\omega\left(\ell_{q\bar{q}}+3/2\right) +m_{q}+m_{\bar{q}}$,
$E_{1}=E_{0}+2\omega$, $E_{2}=E_{1}+2\omega$, $\dots$),
has the following closed form \cite{IJTPGTNO11p179}:
\begin{widetext}
\begin{equation}
S_{ij}(E)\; =\;\delta_{ij}\; -\; 2i\fnd
{
2\lambda^{2}r_{q\bar{q}}
\left\{\dissum{n=0}{\infty}\fnd{g_{i}(n)g_{j}(n)}{E-E_{n}}\right\}
\sqrt{\fnd{\mu_{i}\mu_{j}}{k_{i}k_{j}}}k_{i}k_{j}
j_{\ell_{i}}\left( k_{i}r_{q\bar{q}}\right)
j_{\ell_{j}}\left( k_{j}r_{q\bar{q}}\right)
}
{
1+2i\lambda^{2}r_{q\bar{q}}
\dissum{m=1}{N}
\left\{\dissum{n=0}{\infty}\fnd{\abs{g_{m}(n)}^{2}}{E-E_{n}}\right\}
\mu_{m}k_{m}
j_{\ell_{m}}\left( k_{m}r_{q\bar{q}}\right)
h^{(1)}_{\ell_{m}}\left( k_{m}r_{q\bar{q}}\right)
}
\;\;\; ,
\label{multiS}
\end{equation}
\end{widetext}
where, in the $i$-th channel,
the linear momentum $k_{i}$ and reduced mass $\mu_{i}$
are related to the total invariant mass $E$ of the system,
and to the two meson masses $M_{i1}$ and $M_{i2}$,
through
\begin{eqnarray*}
E & = & \sqrt{k_{i}^{2}+M_{i1}^{2}}+\sqrt{k_{i}^{2}+M_{i2}^{2}}
\;\;\; ,\\ [5pt]
E^{2} & = & 2k_{i}^{2}+M_{i1}^{2}+M_{i2}^{2}+2E\mu_{i}
\;\;\; .
\end{eqnarray*}
$j_{\ell}$ and $h^{(1)}_{\ell}$ represent the spherical Bessel and Hankel
function of the first kind, respectively.

The model parameters representing quark masses ($m_{n}=0.406$ GeV,
$m_{s}=0.508$ GeV, $m_{c}=1.562$ GeV) and the radial
spacings in the bare confinement spectrum ($\omega =0.19$ GeV)
are kept identical to the ones
originally optimized in Ref.~\cite{PRD27p1527}, and also used in
Ref.~\cite{HEPPH0606022}.
Moreover, the parameter $r_{q\bar{q}}$, which stands for
the average radius of $^{3\!}P_{0}$ quark-pair creation,
is identical to the value $r_{sn}=3.2$ GeV$^{-1}$ used in
Ref.~\cite{EPJC22p493}, but scaled with the reduced
quark mass in order to impose flavor symmetry of our equations
\cite{PRD74p037501,MPLA19p1949}, i.e.,
\begin{eqnarray*}
r_{cn} & = &
\fnd{m_{s}\left( m_{c}+m_{n}\right)}{m_{c}\left( m_{s}+m_{n}\right)}r_{sn}
=2.24\;\xrm{GeV}^{-1}
\;\;\; ,\\ [5pt]
r_{cs} & = &
\fnd{m_{n}\left( m_{c}+m_{s}\right)}{m_{c}\left( m_{n}+m_{s}\right)}r_{sn}
=1.88\;\xrm{GeV}^{-1}
\;\;\; .
\end{eqnarray*}
The overall decay coupling constant $\lambda$ is fine-tuned to reproduce
the mass of the now very well established $D_{s0}^{\ast}(2317)$.
Yet, also $\lambda$ turns out to be close
to the values used in the light scalar sector \cite{HEPPH0606022},
owing to the referred flavor-symmetric mass scaling.
This yields the values $\lambda=2.854$
GeV$^{-3/2}$ when only PP channels are included, and $\lambda=2.617$
GeV$^{-3/2}$ with PP as well as VV channels. Note that the former value
of $\lambda$ is fully compatible with the values found for the light
scalars in Ref.~\cite{HEPPH0606022}, which analysis was also restricted
to PP channels. The change in $\lambda$ from the VV channels amounts
to a reduction by less than 10\%.

The channels included in the present work are summarized in
Table~\ref{channels}, their relative couplings to the
$q\bar{q}$ channels in Table~\ref{couplings}.
\begin{table}[htbp]
\begin{center}
\begin{tabular}{||c|c||c|c||}
\hline\hline\multicolumn{2}{||c||}{}&\multicolumn{2}{c||}{}\\[-10pt]
\multicolumn{2}{||c||}{charm-nonstrange} &
\multicolumn{2}{c||}{charm-strange}\\ [3pt]
\hline\hline & & & \\ [-10pt]
Channels&Thresh.\ &Channels&Thresh.\ \\
(waves) & (GeV) & (waves) & (GeV)\\ [3pt]
\hline & & & \\ [-10pt]
$D\pi$ ($S$) &2.004& $DK$ ($S$) &2.363\\
$D\eta$ ($S$) &2.415& $D_{s}\eta$ ($S$) &2.516\\
$D\eta '$ ($S$) &2.825& $D_{s}\eta'$ ($S$) &2.926\\
$D_{s}K$ ($S$) &2.464& & \\
$D^{\ast}\rho$ ($S,D$) &2.784& $D^{\ast}K^{\ast}$ ($S,D$) &2.902\\
$D^{\ast}\omega$ ($S,D$) &2.791& $D_{s}^{\ast}\phi$ ($S,D$) &3.132\\
$D_{s}^{\ast}K^{\ast}$ ($S,D$) &3.006&&\\[3pt]
\hline\hline
\end{tabular}
\end{center}
\caption[]{The various meson-meson channels included in this analysis,
and their threshold energies.}
\label{channels}
\end{table}
\begin{table}[htbp]
\begin{center}
\begin{tabular}{||c||c||}
\hline\hline & \\[-10pt]
charm-nonstrange & charm-strange\\ [3pt]
\hline\hline & \\ [-10pt]
$\sqrt{1/16}$ & $\sqrt{1/12}$\\ [5pt]
$x\sqrt{1/144}-y\sqrt{1/72}$ & $-y\sqrt{1/72}-x\sqrt{1/36}$ \\ [5pt]
$y\sqrt{1/144}+x\sqrt{1/72}$ & $x\sqrt{1/72}-y\sqrt{1/36}$ \\ [5pt]
$\sqrt{1/24}$ & \\ [5pt]
$\sqrt{1/48}\;,\sqrt{5/12}$ & $\sqrt{1/36}\;,\sqrt{5/9}$ \\ [5pt]
$\sqrt{1/144}\;,\sqrt{5/36}$ & $\sqrt{1/72}\;,\sqrt{5/18}$ \\ [5pt]
$\sqrt{1/72}\;,\sqrt{5/18}$ & \\[3pt]
\hline\hline
\end{tabular}
\end{center}
\caption[]{The relative couplings
(four-fermion recombination coefficients \cite{ZPC21p291})
for the various meson-meson channels included in this analysis
(in the same order as in Table~\ref{channels}),
to $J^{PC}=0^{++}$ $c\bar{n}$
and $c\bar{s}$ in $S$ or $D$ waves.
For the relative couplings to higher radial excitations $n$,
one has $g(n)=\sqrt{(n+1)}g(0)/2^{n}$ for $S$
and $g(n)=\sqrt{(2n+5)}g(0)/(5\times 2^{n})$ for $D$ waves.
The symbols $x$ and $y$ stand for $\cos\Theta_{\mbox{\scriptsize PS}}$ and
$\sin\Theta_{\mbox{\scriptsize PS}}$, respectively.}
\label{couplings}
\end{table}

The pseu\-do\-scalar $\eta$-$\eta '$ mixing angle $\Theta_{PS}$ we choose
at the recently found
experimental value $\Theta_{PS}=-13.5^\circ$ \cite{HEPEX0411081} (octet-singlet
basis).
However, we also verify our results for another frequently used value,
i.e., $\Theta_{PS}=-17.3^\circ$ \cite{HEPPH0606022}, which turns out to change
the predictions by only a few MeV. We force the damping of closed scattering
channels with subthreshold form factors, which are a standard tool in modern
multichannel phase-shift analyses:
\begin{equation}
g^2_{i}(n)\;\to\; g^2_{i}(n)\, e^\x{\alpha k_{i}^{2}}
\;\;\;\;\xrm{for}\;\;\;\;
\Re e\, k_{i}^{2}<0
\;\;\; .
\label{damping}
\end{equation}
We choose the value $\alpha=4$ GeV$^{-2}$, which is the same as used
in the analysis of the light scalars \cite{HEPPH0606022}.
Such a suppression, in addition to the one resulting from our kinematically
relativistic Schr\"{o}dinger formalism, can be justified from relativistic
covariance, offshellness, self-energies, and other effects not accounted
for in the present model. These contributions are, of course, very difficult
to rigorously evaluate in our nonperturbative scheme. However, even if we
were to completely switch off subthreshold damping, our $D_{s0}^{\ast}$(2850)
pole would only shift to $2864-i\times15$ MeV, after a readjustment of
$\lambda$ so as to reproduce again the $D_{s0}^{\ast}$(2317) mass.

Having now fixed all parameters in formula~(\ref{multiS}),
we can search our amplitudes for resonance poles.
We predict the first radial excitations of the ${D_{s0}^{\ast}}$(2317)
and $D_0^{\ast}$(2400) to come out as
${D_{s0}^{\ast}}'$(2850) \footnote
{In the one-PP-channel calculations of Ref.~\cite{PRD74p037501} and
Ref.~\cite{MPLA19p1949} (3rd paper), the ${D_{s0}^{\ast}}'$(2850) pole was
estimated at $2923-i57$ MeV and $2928-i20$ MeV, respectively. Here, with a
restriction to the three PP channels only, the pole comes out at $2804-i59.9$
MeV.}
and ${D_{0}^{\ast}}'$(2740), respectively. We furthermore predict the very
broad states ${D_{s0}^{\ast}}$(2780) and ${D_{0}^{\ast}}$(2700), which might
show up in a more pronounced way in production experiments than in elastic
scattering.

Note added in proof: after completion of this work, the BABAR collaboration
posted a preprint \cite{HEPEX0607082} confirming the announcement of the
$D_{sJ}(2860)$ in Ref.~\cite{DsJ2860}. Furthermore, three theoretical papers
on this new state have appeared in the meantime, the first one
\cite{HEPPH0607245} favoring a $3^-$ ($1\,^3\!D_3$) assignment, the second
\cite{HEPPH0608139} a $0^+$ ($2\,^3\!P_0$) like we do, and the third
\cite{HEPPH0609013} admitting either possibility. Clearly, the non-observation
so far of the $D^*\!K$ decay mode, forbidden for a scalar meson, favors the
$0^+$ option, although the $D_s\eta$ mode, not observed either, is allowed in
both the $0^+$ and $3^-$ scenarios. It is also interesting that
Ref.~\cite{HEPPH0607245}, which makes out a case for the $3^-$ assignment,
predicts branching ratios $D_{sJ}(2860)\to D^*\!K/DK\,=\,0.39$ for $3^-$, and
$D_{sJ}(2860)\to D_s\eta/DK\,=\,0.34$ for $0^+$. We predict a value of 0.30 for
the latter branching ratio, if we include all PP+VV channels. Anyhow,
experiment will have the final word on interpreting the $D_{sJ}(2860)$ beyond
any doubt, by observing either $D_s\eta$ or $D^*\!K$. 
\section*{Acknowledgments}
This work was presented at the QNP06 conference in Madrid \cite{Madrid_talk},
unknowing of the simultaneous experimental release at
the CHARM06 conference in Beijing \cite{DsJ2860}.
We are indebted to S.~Tosi for drawing our attention to the brand-new data.
This work was supported in part by the {\it Funda\c{c}\~{a}o para a
Ci\^{e}ncia e a Tecnologia} \/of the {\it Minist\'{e}rio da Ci\^{e}ncia,
Tecnologia e Ensino Superior} \/of Portugal, under contract
POCI/FP/63437/2005.
\enlargethispage*{10pt}

\end{document}